\documentclass[conference]{IEEEtran}
\IEEEoverridecommandlockouts
\usepackage{cite}
\usepackage{amsmath,amssymb,amsfonts}
\usepackage{algorithmic}
\usepackage{graphicx}
\usepackage{textcomp}
\usepackage{xcolor}

\usepackage{multirow}
\usepackage{arydshln}
\usepackage{booktabs}
\usepackage{longtable}
\usepackage{multirow}
\usepackage{caption}
\usepackage{soul}
\usepackage{tikz}
\usetikzlibrary{calc}
\usepackage{fontawesome}

\def\BibTeX{{\rm B\kern-.05em{\sc i\kern-.025em b}\kern-.08em
    T\kern-.1667em\lower.7ex\hbox{E}\kern-.125emX}}

\newcommand{\greencheck}{{\color{black}\checkmark}}

\title{FIDESlib: A Fully-Fledged Open-Source FHE Library for Efficient CKKS on GPUs\\
}
\author{
\IEEEauthorblockN{Carlos Agulló-Domingo} 
\IEEEauthorblockA{\textit{Universidad de Murcia}\\
carlos.a.d@um.es} \\
\IEEEauthorblockN{Kaustubh Shivdikar}
\IEEEauthorblockA{\textit{AMD}\\
kshivdik@amd.com} 
\and 
\IEEEauthorblockN{Óscar Vera-López}
\IEEEauthorblockA{\textit{Universidad de Murcia}\\
oscar.veral@um.es} \\
\IEEEauthorblockN{Rashmi Agrawal}
\IEEEauthorblockA{\textit{Boston University}\\
rashmi23@bu.edu} 
\and 
\IEEEauthorblockN{Seyda Guzelhan}
\IEEEauthorblockA{\textit{Boston University}\\
seyda@bu.edu} \\
\IEEEauthorblockN{David Kaeli}
\IEEEauthorblockA{\textit{Northeastern University}\\
kaeli@ece.neu.edu} 
\and 
\IEEEauthorblockN{Lohit Daksha}
\IEEEauthorblockA{\textit{Boston University}\\
tihol@bu.edu} \\
\IEEEauthorblockN{Ajay Joshi}
\IEEEauthorblockA{\textit{Boston University}\\
joshi@bu.edu} 
\and
\IEEEauthorblockN{Aymane El Jerari}
\IEEEauthorblockA{\textit{Northeastern University}\\
eljerari.a@northeastern.edu} \\
\IEEEauthorblockN{José L. Abellán} 
\IEEEauthorblockA{\textit{Universidad de Murcia}\\
jlabellan@um.es} 
}

\begin{document}

\maketitle

\begin{abstract}
Word-wise Fully Homomorphic Encryption (FHE) schemes, such as CKKS, are gaining significant traction due to their ability to provide post-quantum-resistant, privacy-preserving approximate computing—an especially desirable feature in Machine-Learning-as-a-Service (MLaaS) cloud-computing paradigms.
OpenFHE is a leading CPU-based FHE library with robust CKKS operations, but its server-side performance is not yet sufficient for practical cloud deployment. As GPU computing becomes more common in data centers, many FHE libraries are adding GPU support. However, integrating an efficient GPU backend into OpenFHE is challenging. While OpenFHE uses a Hardware Abstraction Layer (HAL), its flexible architecture sacrifices performance due to the abstraction layers required for multi-scheme and multi-backend compatibility.

In this work, we introduce FIDESlib, the first open-source server-side CKKS GPU library that is fully interoperable with well-established client-side OpenFHE operations. \textit{Unlike other existing open-source GPU libraries}, FIDESlib provides the first implementation featuring heavily optimized GPU kernels for all CKKS primitives, including bootstrapping. Our library also integrates robust benchmarking and testing, ensuring it remains adaptable to further optimization. Furthermore, its software architecture is designed to support extensions to a multi-GPU backend for enhanced acceleration.
Our experiments across various GPU systems and the leading open-source CKKS library to date, Phantom, show that FIDESlib offers superior performance and scalability. For bootstrapping, FIDESlib achieves no less than 70$\times$ speedup over the AVX-optimized OpenFHE implementation.
\end{abstract}

%\begin{IEEEkeywords}
%component, formatting, style, styling, insert.
%\end{IEEEkeywords}

\section{Introduction and Motivation}
\label{sec:intro}

% Machine-Learning-as-a-Service (MLaaS)
Machine-Learning-as-a-Service (MLaaS)~\cite{ribeiro2015mlaas} is a cloud-based powerful tool to simplify the development, deployment, and management of machine learning models. Nowadays, major cloud providers enable robust ML platforms such as Amazon SageMaker~\cite{mishra2019machine} for data preparation, training, and deployment; Google Vertex AI~\cite{ggp2023} for AutoML and custom training; and Azure Machine Learning~\cite{korner2020mastering} for automated end-to-end workflows.
% Data breaches and quantum computers
While MLaaS platforms are gaining traction, data breaches pose significant risks, leading to millions in losses~\cite{ibmDataBreachReport}. This threat is further amplified by the anticipated rise of quantum computers, which are expected to possess encryption-breaking capabilities within the next decade~\cite{lloyd-jones2024quantum}. 
Fortunately, lattice-based cryptography, endorsed by NIST, offers a solution, with post-quantum-resistant standards such as FIPS 203~\cite{nistFips203} and FIPS 204~\cite{nistFips204} recently announced~\cite{nistAnnounce}.

% CKKS
CKKS~\cite{ckks2017} is a modern, Ring-Learning-With-Errors (RLWE) lattice-based cryptographic scheme that is emerging as the preferred Fully Homomorphic Encryption (FHE) technique for privacy-preserving approximate computing (PPaC), thereby potentially addressing the critical data breach challenges in MLaaS-enabled cloud platforms.
% Computational challenges
However, 
%the significantly larger size of CKKS-encrypted data (with ciphertexts typically being hundreds of times larger than the original plaintext) and 
the large computational overhead of CKKS incurs a 2--5 orders of magnitude slowdown~\cite{heaan_slowdown}, hindering its practical deployment in many of these cloud services, such as real-time ML inference.

% OpenFHE
OpenFHE~\cite{ckks2017} is the leading CPU-based FHE library, extensively supported by a large, multidisciplinary research community. As a result, OpenFHE incorporates robust and rigorously validated client-side CKKS operations (e.g., encryption/descryption) and server-side CKKS operations (e.g., modular arithmetic and bootstrapping). 
However, despite its AVX-optimized CPU backend~\cite{boemer2021intelhexlacceleratinghomomorphic}, the server-side CKKS operations remain far from achieving the level of performance required for practical deployment in this class of cloud services.
Although FPGA and ASIC CKKS accelerators~\cite{survey_impls} could achieve the desired practicality, their widespread adoption as established compute nodes in data centers faces significant challenges. These include the high integration costs within existing and mature MLaaS ecosystems, making it difficult for them to become a reality.

Given the growing presence of GPU-based computing nodes in high-performance data centers~\cite{gpus-datacenters}, accelerating modern CKKS-powered cloud services by fully leveraging the capabilities of GPU platforms is becoming increasingly common. 
Consequently, most state-of-the-art CKKS libraries are already utilizing GPUs~\cite{yang2024phantom,kim2024cheddar,heaan2022,heongpu2024,jung2021over100,troynova2023,wang2023hebooster,fan2023tensorfhe,desilo2023liberatefhe}. However, while adding a GPU backend to OpenFHE would mark significant progress, its multi-scheme and multi-backend general software architecture inherently sacrifices cutting-edge performance in favor of generality, due to the many abstraction layers needed for such flexibility.

%The GPU architecture is well-suited for accelerating the SIMD nature of CKKS computations, achieving short- and mid-term solutions for privacy-preserving computing systems~\cite{gilad2016cryptonets, juvekar2018gazelle} and applications such as encrypted logistic regression training~\cite{han2018logreg}, encrypted Convolutional Neural Network (CNN) inference~\cite{lee2022multiplexedpacking}, and encrypted large language model (LLM) inference~\cite{zama}. 

% GPU CKKS libraries
State-of-the-art GPU libraries that efficiently support all CKKS functionality are often kept private for internal use or industry applications~\cite{heaan2022,fan2023tensorfhe,kim2024cheddar}. This lack of open access limits collaboration across the broader multi-disciplinary research community, hindering the development of new algorithmic optimizations aligned with the rapid advancements in GPU architectures. It also restricts the adoption of CKKS in MLaaS scenarios and prevents general users from benefiting from these optimizations with privacy-preserving computation guarantees. To address these challenges, democratizing CKKS optimization research for modern GPUs is essential.

% FIDESlib
To achieve this important goal, we propose FIDESlib, the first open-source CKKS GPU library that is \textit{feature-complete}, including bootstrapping, and overcomes the limitations of existing open-source counterparts, delivering the highest performance.
To facilitate collaborative development and further optimization of our library, our design strategy for FIDESlib is threefold. First, we leverage the well-established and robust OpenFHE library for all client-side operations. Second, we implement all CKKS server-side operations, adding exhaustive testing capabilities (unit and integration tests) for complete functional validation against OpenFHE’s server operations, using the GoogleTest framework~\cite{googletest}. Third, we enable extensive microbenchmarking and benchmarking capability by utilizing the Google Benchmark framework~\cite{googlebench}.

Our key contributions are as follows.
\begin{itemize}
    \item We propose FIDESlib\footnote{We plan to release FIDESlib after completion of the review process.}, the first open-source implementation of CKKS with full functionality, including bootstrapping, specifically optimized to maximize compute throughput, on-chip data reuse, and memory bandwidth utilization for a GPU backend.
    \item FIDESlib is the first GPU library that provides full interoperability with the standard OpenFHE framework, enabling the implementation of server-side CKKS operations while delegating encoding and encryption tasks to OpenFHE-based clients.
    \item Our experiments across various GPU platforms show that FIDESlib outperforms Phantom~\cite{yang2024phantom}, the leading open-source CKKS library, in performance and scalability for all operations supported by Phantom. For CKKS operations not supported by Phantom, such as bootstrapping, FIDESlib achieves up to 74$\times$ speedup over the best-performing AVX-optimized OpenFHE implementation.
\end{itemize}

\newcommand{\plus}{\raisebox{.1\height}{\scalebox{.9}{+}}}

\section{Background}

\label{sec:background}

% note: assuming the long versions of the abbreviations HE and FHE exist in the introduction
% There is a way to use acronyms throughout, but the first use will spell them out.  I think it is the acronym package - maybe use in the future.
\subsection{CKKS Optimization Techniques}

HE enables operations on encrypted data without having access to the secret key. Following Gentry's blueprint~\cite{gentry2009}, several FHE schemes have been developed to support exact arithmetic on integers (BGV~\cite{bgv2014} and BFV~\cite{bfv}), approximate arithmetic on fixed-point numbers (CKKS~\cite{ckks2017}), and Boolean gates (TFHE~\cite{chillotti2020tfhe} and FHEW~\cite{ducas2015fhew}). %Although the theoretical guarantees provide a holy grail for data security, the resulting ciphertext size and computational overhead prevent practical deployment of FHE~\cite{gong2025security} in many scenarios. \\
Given CKKS's potential to enable PPaC in MLaaS cloud environments, we have implemented and optimized this scheme in FIDESlib, targeting a GPU backend.
Table~\ref{tab:primitives} lists the available primitives of the CKKS cryptosystem, while Table~\ref{tab:ckks_params} presents the main CKKS parameters and their notation used in this work.
For a detailed description of the CKKS scheme, refer to~\cite{cheon2017homomorphic}.

\begin{table}[t!]
\begin{tabular}{@{}lll@{}}
 \toprule
 \textbf{Operation} & \textbf{Output} & \textbf{Description}  \\
 \hline \hline 
ScalarAdd($|| x ||, c)$ & $|| x + c ||$ & Adding a constant to a ciphertext. \\
 \hline
 PtAdd($|| x ||, y)$ & $|| x \plus y ||$ & Addition of a plaintext with a ciphertext.  \\
 \hline
 HAdd($|| x ||, || y ||)$ & $|| x \plus y ||$ & Addition of two ciphertexts.   \\
  \hline
 ScalarMult($|| x ||, c)$ & $|| x \odot c ||$ & Multiplying a ciphertext by a constant.   \\
 \hline
  PtMult($|| x ||, y)$ & $|| x \odot y ||$ & Multiplying a ciphertext by a plaintext.   \\
  \hline
 HMult($|| x ||, || y ||)$ & $|| x \odot y ||$ & Multiplication of two ciphertexts. \\ 
 \hline
  Rescale($|| x ||_{C_L})$ & $|| x ||_{C_{L-1}}$ & Rescale after multiplication.  \\
 \hline
 Conjugate($|| x ||)$ & $|| \bar{x} ||$ & Conjugation of the underlying message. \\  
 \hline
  HRotate($|| x ||, k)$ & $|| \phi_k(x) ||$ & Rotation of the underlying message by k.  \\

 \bottomrule
\end{tabular}
\caption{CKKS RNS primitives.}
\label{tab:primitives}
\end{table}

\begin{table}[t!]
\centering
\begin{tabular}{c l} 
    \toprule
    \textbf{Param} & \textbf{Description} \\ [0.5ex] 
    \hline
    \hline
    $N$ & Polynomial degree-bound\\%. Length of the plaintext polynomial \\
    $n$ & Length of the message. $n \leq \frac{N}{2}$ \\
    $Q$ & Polynomial modulus \\
    $L$ & Multiplicative depth before\\ & bootstrapping is needed\\
    $\mathcal{B}$ & The set $\{q_0, q_1, \dots, q_L\}$ of prime factors of $Q$\\% = \prod_{i=0}^{L}q_i$\\
    $\ell + 1$ & Current number of limbs\\ $\ell$ & remaining multiplicative depth\\%Current number of limbs in a ciphertext \\
    $\mathsf{dnum}$ & Number of digits in the switching key \\
    %$\alpha$ & Number of limbs that comprise a single digit \\ 
 %& in the key-switching decomposition $\alpha = \lceil \frac{L + 1}{\mathrm{dnum}} \rceil $ \\
    $P$ & Product of extension limbs added for \\
     & raised modulus. \\
   %  Total extension limbs $= \alpha + 1$ \\
    %  & linear transform \\
   $\Delta$ & scaling applied during encoding. $q_i \approx \Delta$ \\
   %$\mathbf{m}$ & A message vector of $n$ slots \\
%   $\dbm$ & Ciphertext encrypting a message \\
%   $\amp$ & A randomly sampled polynomial from message $\mathbf{m}$ \\
   $x$ & a polynomial \\
  % $P_m$ & Polynomial encrypting message $m$ \\
   $[x]_{q_{i}}$ & $q_i$-limb of $x$ \\
   $\mathbf{ksk}$ & Evaluation/key-switching key \\
   NTT & Number Theoretic Transform, takes a coefficient \\
   & vector and outputs an evaluation vector \\
   iNTT & Inverse Number Theoretic Transform, takes an \\
   & evaluation vector and outputs a coefficient vector \\
   $\psi$ & $2n$-th root of unity under a prime $q_i$, \\
   & typically precomputed for (i)NTT evaluation. \\
   %$\mathbf{ksk}_{rot}^{(r)}$ & Evaluation key for \textit{HE-Rotate} block with\\
   % & $(r)$ rotations \\
    \bottomrule
\end{tabular}
%\vspace{-2.13em}
\caption{CKKS Notation and descriptions.}
\label{tab:ckks_params}
\end{table}

%
% Moreover, the number of multiplicative operations on ciphertexts is limited by the noise accumulation, and the bootstrapping operation is needed to refresh the number of multiplications available to avoid corrupting the underlying message. This noise accumulation is called level of the ciphertext. The message is scaled before the homomorphic operations to reduce the impact of noise. As each multiplication results in a squared scaling factor, the Rescale operation is used to down-scale the message back to the original factor. Rotation and multiplication operations require key-switching operations to recover the correct encryption key by rotating the same amount and removing the scaled factors, respectively. Although they provide versatility, their high computational cost and memory requirements create a bottleneck. \\

In contrast to BFV and BGV, CKKS treats the fresh encryption error as part of the noise from fixed-point operations~\cite{ckks2017}. The most well-known CKKS libraries (HElib~\cite{helib}, HeaNN~\cite{ckks2017}, SEAL~\cite{seal}, OpenFHE~\cite{openfhe}) provide efficient implementations of FHE on CPU platforms. 
Several algorithmic optimizations exist to reduce computation costs and enhance CKKS precision, which can also be applied to a GPU backend. Below, we discuss the most significant of these optimizations.

The large size of the ciphertext modulus can be resolved by using the Residue Number System (RNS) technique based on the Chinese Remainder Theorem~\cite{cheon2019rns}. Precision of rescaling under RNS can be improved by carefully tracking the scaling factors at each level~\cite{kimpol2022reduced}. 

Key-Switching is needed by HMult and HRotate operations to keep the encrypting key of the result the same as the input by employing a ``switching key''. Hybrid key switching proposes the idea of decomposing the switching keys into $dnum$ chunks so that the noise handling is easier in each small partition \cite{hk2020}. This comes with a trade-off of increased computation and key-size. 
% Hybrid key switching is proposed for a better operational cost versus ciphertext modulus consumption trade-off, reducing noise during key switching more efficiently~\cite{hk2020}.

The number of multiplicative operations on a ciphertext is limited by the amount of accumulated noise. The bootstrapping operation is used to refresh the ciphertext~\cite{chkks2018} noise and allow for further computation. Algorithmic improvements to bootstrapping focus mainly on linear transforms~\cite{hs2018bts, cheon2018faster} or polynomial evaluation~\cite{lee2021bts}, or both~\cite{hk2020, bossuat2021, bossuat2022}. The linear transforms can be improved by exploiting matrix sparsity~\cite{chen2019improved} and applying decomposition methods~\cite{cheon2018faster} to the original DFT matrix. Even with these optimizations, bootstrapping consumes more than half of the runtime of a typical FHE-enabled application when run on a GPU~\cite{jung2021over100}.

\subsection{The Memory Bottleneck and GPUs}

As explored in many prior studies~\cite{mad, ark, dfhenca}, FHE workloads are heavily memory-bound. This is primarily due to the combination of low arithmetic intensity operations, large ciphertext and key size, and lack of advances in memory bandwidth and latency in modern compute platforms~\cite{book}. As a result, FHE workloads require new acceleration strategies. Computations performed in the ciphertext domain are typically 2 to 5 orders of magnitude slower than their plaintext counterparts~\cite{heaan_slowdown}. This degree of slowdown is due to the use of large polynomials that encode plaintext data used in ciphertext computations. Large polynomial parameters need to be used to maintain adequate levels of security, as well as to ensure noise growth does not impact the underlying data. For example, to achieve 128-bit security, FHE operations require using polynomials of a degree up to $2^{17}$, with the size of each coefficient being up to $2200$ bits~\cite{shivdikar2023gme}.

Unlike CPU-based FHE implementations, which are constrained by the bandwidth of DDR memory, GPUs can leverage GDDR or High Bandwidth Memory (HBM) modules, helping to alleviate the memory bottleneck. While modern DDR5-based systems can achieve a peak memory bandwidth of around 100GB/s, consumer GPUs like the RTX 4090 used in our experiments exceed 1 TB/s of memory bandwidth, significantly accelerating CKKS primitives. Additionally, the on-chip memory capacity in GPUs plays a crucial role in data reuse effectiveness. When properly utilized, it reduces costly off-chip memory accesses. For instance, the RTX 4090 includes a large 72MB L2 cache, which will be essential for further acceleration in FIDESlib. 
Although CKKS primitives are not typically compute-intensive, their performance on GPUs can be significantly affected by modular arithmetic operations on integers (the key operations in CKKS), as current GPU architectures lack efficient hardware support for modulo operations. To demonstrate the efficiency of our FIDESlib library across different GPU platform capabilities, we will test four distinct GPU platforms in Section~\ref{sec:evaluation} (listed in Table~\ref{tab:gpu_specs}).

% TOPS
% Mmeory hierarhcy

% \begin{table}
% \centering
% \begin{tabular}{@{}lccc@{}}
% \toprule
% \textbf{Memory}   & \textbf{Cached?} & \textbf{Access}         & \textbf{Lifetime} \\ \midrule
% Register & N/A     & Read and Write & Thread   \\
% Global   & \(\times\)      & Read and Write & Kernel   \\
% Shared   & N/A     & Read and Write & Block    \\
% Texture  & \(\checkmark\)     & Read only      & Kernel   \\
% Constant & \(\checkmark\)     & Read only      & Kernel   \\ \bottomrule
% \end{tabular}
% \caption{Various GPU memory spaces offered by the CUDA programming model}
% \label{tab:cuda_memory}
% \end{table}

%% Shift to Intro/Motivation sections
%First implementation with bootstrapping: \cite{jung2021over100}. Polynomial multiplication on GPUs \cite{shivdikar2022polymult}, GME \cite{shivdikar2023gme}, accelerating NTT \cite{kim2020gpuntt}, double scaling \cite{agrawal2023doublescaling}.
% Libraries: Phantom \cite{yang2024phantom}, Cheddar \cite{kim2024cheddar}, Chameleon \cite{wang2024chameleon}. 

\section{FIDESlib}
\label{sec:fideslib}

%Next, we provide an overview of the design of FIDESlib, including its software architecture and implementation details. We also cover the key algorithms and optimizations used in our implementation.

\subsection{Functionality Overview}

% FIDESlib implements all every server-side CKKS operation on GPU (HAdd, HMult, etc), while leaving client operations (KeyGen, Encrypt, Decrypt, etc) to be executed by the OpenFHE library. Figure~\ref{fig:functionality} illustrates how functionality is split amongst both libraries.

% To the authors' knowledge, FIDESlib is the first open source implementation of the bootstrapping procedure for CKKS on GPU. Furthermore, FIDESlib implements ScalarAdd, ScalarMult and HSquare, which may be considered optimized versions of PtAdd, PtMult and HMult for restricted input spaces, rarely provided by other libraries. Finally, the HoistedRotation optimized routine is also implemented, reducing the cost of performing several rotations of a single ciphertext.

\begin{figure}[htbp]
\centerline{
    \includegraphics[width=0.5\textwidth, clip=true, trim=-5 0 0 0mm]{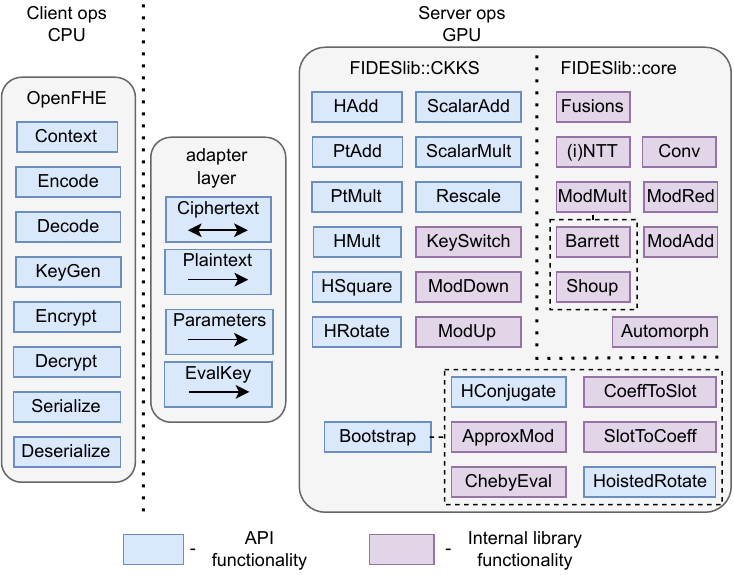}
}
\caption{High-Level Functionality Overview in FIDESlib.}
\label{fig:functionality}
\end{figure}

FIDESlib implements all server-side CKKS operations (e.g., HAdd, HMult) on the GPU, while client-side operations (e.g., KeyGen, Encrypt, Decrypt) are handled by the OpenFHE library. Figure~\ref{fig:functionality} shows how functionality is divided into the two libraries. To the best of our  knowledge, FIDESlib includes the first open-source GPU implementation of the CKKS bootstrapping procedure. FIDESlib also offers optimized routines such as ScalarAdd, ScalarMult and HSquare, which are more efficient versions of PtAdd, PtMult, and HMult for inputs with repetitive data (features rarely available in other libraries). Additionally, the HoistedRotation routine \cite{hs2018bts} is implemented to reduce the cost of multiple ciphertext rotations.

\subsection{OpenFHE interoperation}
%FIDESlib currently supports one of the four CKKS modes of operation provided by OpenFHE~\cite{openfhe}, namely \textit{FIXEDMANUAL}. Although this mode offers the lowest precision, it provides the best performance. The modifications required to support other modes, such as \textit{FLEXIBLEAUTOEXT}, are not algorithmically complex and will be incorporated into FIDESlib in the near future.

OpenFHE adheres to the \textit{HomomorphicEncryption.org} post-quantum security standards for homomorphic encryption. As the security of the library depends solely on the correctness of client-side operations, users of the FIDESlib library benefit from the same security guarantees. However, the OpenFHE and FIDESlib source codes are decoupled by implementing a thin adapter layer. Loose coupling offers several benefits:
\begin{itemize}
\item {\bf Changes to OpenFHE}:
Only minimal changes are made to the OpenFHE codebase, ensuring that new versions of the OpenFHE library are unlikely to break compatibility. In our repository, the instructions for compiling a compatible version of OpenFHE for FIDESlib include applying a small Git patch to any OpenFHE release.
\item {\bf Freedom and Simplicity of Implementation}: 
The generic multi-backend design of OpenFHE's implementation of multi-FHE schemes makes developing a performant GPU acceleration backend challenging. By adopting our own code structure and focusing on the CKKS scheme, we reduce code complexity while maintaining the flexibility required for further optimization.
\item {\bf Code Repurposing}:
Since the core functionality of the library is not tied to OpenFHE's code, FIDESlib can be easily adapted as a GPU backend for other projects.
\end{itemize}

Internally, the adapter layer transfers data between OpenFHE's objects and simplified data structures that retain essential data and metadata fields. These simplified data structures are then passed to methods within FIDESlib's classes, such as \textit{CKKS::Parameters}, \textit{CKKS::Ciphertext}, \textit{CKKS::Plaintext}, and \textit{CKKS::KeySwitchingKey}, which store the data in GPU memory. For the \textit{CKKS::Ciphertext} class, the reverse process is also supported, transferring the data, along with static noise estimation data, back to an OpenFHE ciphertext object for decryption.

\subsection{Software design}

The FIDESlib codebase is organized into four namespaces:
\begin{enumerate}
\item \textbf{FIDESlib}: core functionality and utility code to support power of two polynomial ring arithmetic under word-sized moduli on GPU.
\item \textbf{FIDESlib::CKKS}: implementation of the CKKS's functionality, including GPU kernels, a composition-based hierarchy of GPU-memory-managing classes, and the CKKS crypto-context class.
\item \textbf{FIDESlib::test}: a Google Test-based test suite comprising both unit and integration tests.
\item \textbf{FIDESlib::bench}: a Google Bench-based benchmark suite for testing the library's functionality and performance with synthetic and real workloads.
\end{enumerate}

In the remainder of this section, we focus on the \textbf{FIDESlib} and \textbf{FIDESlib::CKKS} namespaces. For brevity, the \textbf{FIDESlib::CKKS} prefix will be omitted.

\begin{figure}[t!]
\centerline{
    \includegraphics[width=0.4\textwidth, clip=true, trim=0 4 0 4mm]{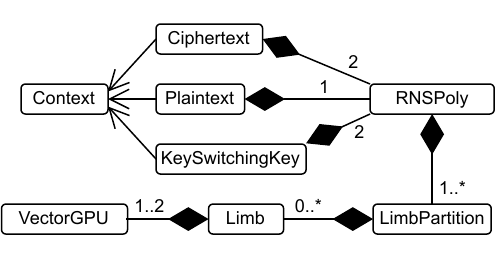}
}
\caption{Data class diagram.}
\label{fig:data}
\end{figure}

The organization of data within the library is illustrated in Figure~\ref{fig:data}, which is discussed below.

\textit{Ciphertext}, \textit{Plaintext}, and \textit{KeySwitchingKey} are composed of a number of \textit{RNSPoly} objects. Each \textit{RNSPoly} represents an $N$ degree polynomial, as determined by its generating \textit{Context}. The polynomial is defined under a specific integer modulus and decomposed into a base $\mathcal{B} = {q_0, q_1, ..., q_l}$, as specified by the Remainder Number System.

Each \textit{RNSPoly} object is composed of one or more \textit{LimbPartition} objects, which represent the portion of a polynomial stored in a specific device. As the current FIDESlib version supports single-GPU operations (with multi-GPU support in development), each \textit{RNSPoly} object can be assumed to contain one \textit{LimbPartition}, with all \textit{LimbPartition} instances assigned to the same GPU.

Each \textit{LimbPartition} contains one or several \textit{stacks} of \textit{Limb} objects, which represent a polynomial under a specific base modulus $q_i \in \mathcal{B}$. The \textit{LimbPartition} class accesses metadata arrays specific to each \textit{Limb stack} type it stores and provides methods for managing them. These metadata stacks are defined during \textit{Context} creation to simplify precomputation and GPU execution setup.
%, contrasting with OpenFHE's more dynamic approach to configuring polynomials in RNS bases.

Finally, a \textit{Limb} instance contains a \textit{VectorGPU} object that manages a contiguous array of memory and an optional second \textit{VectorGPU} object for $(i)NTT$ computation. The \textit{Limb} class is templated to support both $64$-bit and $32$-bit integers, although the current version only verifies and utilizes $64$-bit integers.

\subsection{Memory management}

%As opposed to most proposals, 
FIDESlib employs a stack of arrays approach instead of using a flattened 2D array of coefficients to represent an RNS polynomial. The main benefit of this approach is its finer grain memory management, which can reduce internal fragmentation. Additionally, we argue that it simplifies programming.

The \textit{VectorGPU} class simplifies memory management using the \textbf{RAII} (Resource Acquisition Is Initialization) pattern, asynchronously allocating device memory at construction and asynchronously freeing it at destruction—i.e., when the object goes out of scope—removing the need for explicit device memory management by the programmer.
Asynchronous device memory management is achieved through the CUDA Stream Ordered Memory Allocator, a.k.a. the default \textit{cudaMemPool\_t}.

%Still, the performance question remains regarding the efficiency of this approach, particularly with the increased management cost of multiple small arrays compared to a single, larger data array. 
%While the library programmer might appreciate the easier use of the stack-of-arrays representation, we also provide the capability of creating unmanaged \textit{VectorGPU} objects. These objects are instantiated with the device data pointer explicitly passed to the constructor, transferring the responsibility for correct memory management to a higher class in the hierarchy.
Although, the stack-of-arrays representation simplifies usage for library programmers, we also offer the option to create unmanaged \textit{VectorGPU} objects. These are instantiated by explicitly passing the device data pointer to the constructor, shifting memory management responsibility to a higher-level class.
We observe that for short-lived or constant-sized \textit{RNSPoly} objects, creating a flattened 2D data array while simulating a stack of arrays through the use of unmanaged \textit{VectorGPU}s can be preferable. In contrast, long-lived and variably-sized \textit{RNSPoly} objects are more easily and memory efficiently managed as true stacks of arrays. Thus, FIDESlib employs both approaches.

\subsection{Precomputation}

Many values are precomputed for the CKKS scheme before server operations are executed, reducing the computational complexity of many operations. For simplicity, most of this precomputation is done during \textit{Context} creation time. 
Next, we discuss these precomputed values and their implications for algorithm implementation.

In OpenFHE, multiple cryptocontexts can be created simultaneously. However, two limitations of the GPU platform make this impractical. The first limitation is the limited size of device memory. The second is that constant memory fields must be defined at compile-time and are restricted to an aggregated size of 64 KB. Running a scheme with different parameter sets concurrently becomes challenging to implement and impractical for a GPU backend.

Consequently, FIDESlib treats Contexts and precomputation results using the singleton software design pattern. This approach allows for the use of constant memory and enables precomputation values to be declared globally. While this may seem unorthodox, it simplifies kernel calls, as the precomputed values do not need to be passed as kernel parameters.
Note that application developers can bypass this library limitation by managing different parameter sets within separate processes, where FIDESlib is linked statically. However, this approach does not resolve the issue of device memory usage.

Finally, values to be used by every thread in a \textit{warp} simultaneously are declared \textit{\_\_constant\_\_}, as the hardware efficiently broadcasts this value to all threads. For values that are worth precomputing but not necessarily accessed by all threads at once, they are declared \textit{\_\_global\_\_}.

\subsection{Algorithms and optimizations}

In this subsection, we describe the core algorithms and optimizations implemented in FIDESlib.

\subsubsection{Stream Handling and Limb Batching}

Computational kernels in the CKKS scheme can be categorized based on their data dependencies: 
\begin{itemize} 
\item \textbf{No dependencies}: These include elementwise operations such as modular multiplication, modular addition, and modulus switching. 
\item \textbf{Dependencies amongst elements within a limb}: Operations like NTT, iNTT, and automorphism are dependent on the data within a single limb. 
\item \textbf{Dependencies amongst elements in the same index of different limbs}: Operations such as ModUp, ModDown, and Rescale involve dependencies across elements in different limbs at the same index. 
\end{itemize}

Given these dependencies, the minimum working set size for complex CKKS operations, such as HMult, corresponds to an entire ciphertext plus a key switching key, which is typically very large (e.g. 120 MB) and challenging to retain on-chip. However, the GPU’s L2 cache hit rate can be improved by performing consecutive operations on a subset of a ciphertext’s limbs before the data is evicted to main memory. This approach has been shown to deliver significant performance gains \cite{mad}. Nonetheless, working with only a subset of a ciphertext’s limbs may lead to under-utilization of the GPU’s compute throughput.

To tackle this problem, FIDESlib divides operations without dependencies across different limbs into multiple kernels, with each kernel processing one or more limbs independently. These independent kernels run asynchronously in separate CUDA streams, maximizing compute utilization. Using kernels with smaller memory footprints enhances temporal locality in data access, thereby improving L2 cache hit rates.

While assigning a single limb per kernel improves locality, the CPU's kernel launch overhead becomes a bottleneck on faster GPUs, which complete tasks quicker than the CPU can schedule. To mitigate this, FIDESlib introduces ``limb-batching'' controlled by a configurable \textit{Context} parameter, to reduce CPU overhead. 

\subsubsection{Modular arithmetic}
Since integer operations on limb elements must be performed modulo a specific set of prime numbers $\{p_i\}_0^{L+K}$ selected at runtime, it is essential to optimize the modulo operation because it is not natively supported by GPUs. Note that using the modulo operator \% naively in CUDA code leads to complex compiler generated routines, which require numerous assembly instructions when the modulo operand is not known at compile time.

Modular addition and subtraction are relatively simple to implement, as their results fall within the ranges $[0, 2p-1)$ and $(-p, p)$, respectively, and can be easily adjusted back to $[0, p)$. In contrast, reducing the results of multiplication operations is significantly more challenging, as they fall within $[0, p^2 - p)$.

Several optimized modular reduction routines have been developed, and Table~\ref{tab:mod_red} compares some of the most commonly used techniques in the literature. Barrett and Montgomery reduction methods improve the efficiency of modular reduction by precomputing auxiliary values dependent on the modulus. The Shoup technique, while more computationally efficient, requires precomputing a constant based on both the modulus and one of the operands.
\begin{table}[]
\centering
\begin{tabular}{@{}lccc@{}}
\toprule
\textbf{Method}           & \textbf{Multiplications} & \textbf{Output}      & \textbf{Observation}                                                                                   \\ \hline
\begin{tabular}[c]{@{}l@{}}Montgomery \\ reduction\end{tabular}      & 1 (wide) + 1 (low)       & \multirow{2}{*}{$[0, 2p)$} & \multirow{2}{*}{\begin{tabular}[c]{@{}l@{}}Requires inputs\\to be in Montgomery\\format\end{tabular}} \\ \cline{1-2}
\begin{tabular}[c]{@{}l@{}}Montgomery \\ multiplication\end{tabular} & 2 (wide) + 1 (low)       &                            &                                                                                                         \\ \hline
\begin{tabular}[c]{@{}l@{}}Shoup \\ multiplication\end{tabular}      & 1 (wide) + 2 (low)       & $[0, 2p)$                                  & \begin{tabular}[c]{@{}l@{}}Precomputation \\depends on one\\ of the inputs\end{tabular}  \\ \hline
\begin{tabular}[c]{@{}l@{}}Barrett \\ reduction\end{tabular}        & 1 (wide) + 1 (low)       & \multirow{2}{*}{$[0, 2p)$} & \multirow{2}{*}{}                                                                                      \\ \cline{1-2}
\begin{tabular}[c]{@{}l@{}}Barrett \\ multiplication\end{tabular}    & 2 (wide) + 1 (low)       &                            &                                                                                                         \\ \bottomrule
\end{tabular}
\caption{Comparison of fast modular reduction methods. Montgomery, Shoup, and improved Barrett\cite{shivdikar2022polymult}. Wide multiplications ($64$-bit $\times$ $64$-bit $\xrightarrow{}$ $128$-bit) are considerably more expensive than low multiplications ($64$-bit $\times$ $64$-bit $\xrightarrow{}$ $64$-bit). } 
\label{tab:mod_red}
\end{table}
FIDESlib implements the improved Barrett reduction \cite{shivdikar2022polymult}, which offers the same computational efficiency as the Montgomery technique but does not require a specific encoding of the elements. Additionally, FIDESlib leverages Shoup’s modular multiplication where appropriate, further accelerating certain modulo operations.
\subsubsection{The base conversion kernel}
The main algorithmic component of the ModDown, ModUp, and Rescale operations is the fast base conversion algorithm. Rescale is a special case where the CRT base is modified by a single prime number, so it is optimized separately. ModDown and ModUp, on the other hand, are used in the (Hybrid) KeySwitching procedure during HMult and HRotate.

Computationally, the fast base conversion algorithm ($Conv_{\mathcal{B'} \xrightarrow{} \mathcal{B}}([x(X)]_{\mathcal{B'}}) = [x(X)]_{\mathcal{B}}$) can be thought of as a modular matrix-vector multiplication, preceded by a limb-wise scaling of the coefficients, as shown by Equation~\ref{eq:conv}. The same computation is applied to every coefficient of the limb, making it a matrix-matrix product.

% \begin{equation}
% \begin{gathered}
%  [x(X)]_{\mathcal{B}} = 
%  \begin{pmatrix}
%      {x(X)}^{(0)} \\
%      {x(X)}^{(1)} \\
%      ... \\
%      {x(X)}^{({K - 1})}
%  \end{pmatrix}
%   = \\
%    \begin{pmatrix}
%      [\hat{q}_{0}]_{p_0} & [\hat{q}_{1}]_{p_0}& ... & [\hat{q}_{L - 1}]_{p_0}\\
%      [\hat{q}_{0}]_{p_1} & [\hat{q}_{1}]_{p_1}& ... & [\hat{q}_{L-1}]_{p_1}\\
%      ... & ... &  &...\\
%      [\hat{q}_{0}]_{p_{K-1}} & [\hat{q}_{1}]_{p_{K-1}}& ... & [\hat{q}_{L-1}]_{p_{K-1}}
%  \end{pmatrix}
%  ·
%  \begin{pmatrix}
%     \hat{q}_{0}^{-1} · {x(X)}^{(0)} \\
%     \hat{q}_{1}^{-1} · {x(X)}^{(1)} \\
%      ... \\
%     \hat{q}_{L - 1}^{-1} · {x(X)}^{({L - 1})}
%  \end{pmatrix}
% \end{gathered}
% \label{eq:conv}
% \end{equation}

\begin{equation}
\footnotesize 
\begin{pmatrix}
    {x(X)}^{(0)} \\
    {x(X)}^{(1)} \\
    \vdots \\
    {x(X)}^{(K-1)}
\end{pmatrix}
=
\begin{pmatrix}
    [\hat{q}_0]_{p_0} \!\!\!\! & \cdots \!\!\!\! & [\hat{q}_{L-1}]_{p_0} \\
    [\hat{q}_0]_{p_1} \!\!\!\! & \cdots \!\!\!\! & [\hat{q}_{L-1}]_{p_1} \\
    \vdots \!\!\!\! & \ddots \!\!\!\! & \vdots \\
    [\hat{q}_0]_{p_{K-1}} \!\!\!\! & \cdots \!\!\!\! & [\hat{q}_{L-1}]_{p_{K-1}}
\end{pmatrix}
\cdot
\begin{pmatrix}
    \hat{q}_0^{-1} \cdot {x(X)}^{(0)} \\
    \hat{q}_1^{-1} \cdot {x(X)}^{(1)} \\
    \vdots \\
    \hat{q}_{L-1}^{-1} \cdot {x(X)}^{(L-1)}
\end{pmatrix}
\label{eq:conv}
\end{equation}

As digit decomposition is applied before base conversion in ModUp, the original base, with $(L+1)/dnum$ elements, becomes up to $dnum$ times smaller than the output base, which has $l+1$ elements. Typically, values for $L/dnum$ are small, so we cache the result of the initial scaling applied to a subset of the limb coefficients in shared memory. We then accumulate each dot product and write the results back to main memory. Two GPU threads are used per vector-matrix operation, maximizing compute resource utilization and ensuring optimal performance.

The kernel thus utilizes $4 \cdot L / dnum$ shared memory bytes per thread. Each shared memory element being reused $l + 1$ times, the number of remaining limbs that remain as multiplicative levels are used. 

To optimize performance, partial dot product results are accumulated as a 128-bit integer and are only reduced modulo $p_i$ before being written back to global memory, saving $l$ reductions per output element. Nonetheless, the base conversion kernel remains compute-bound, especially with large parameter sets.

\subsubsection{(i)NTT}
The (i)NTT kernel enables efficient transformation of polynomials between the coefficient and evaluation spaces. Additions can be performed in either representation, while polynomial multiplication is efficiently carried out in the evaluation space. Other transformations, such as base conversion, must be executed in the coefficient representation.

Internally, the NTT algorithm is implemented as a negacyclic convolution of polynomial coefficients, similar to a Discrete Fourier Transform (DFT) calculation but performed over an integer ring rather than complex numbers. Consequently, the NTT can be implemented using a modified version of the FFT algorithm. Given that the ring dimension $N$ is defined as a power of two, FIDESlib employs the Radix-2 FFT scheme. While larger radices have been proposed (Radix-8~\cite{yang2024phantom}), the Radix-2 algorithm minimizes computational complexity, which we found to be the primary bottleneck.

\begin{figure}[htbp]
\centerline{
    \includegraphics[width=0.45\textwidth, clip=true, trim=0 0 0 0mm]{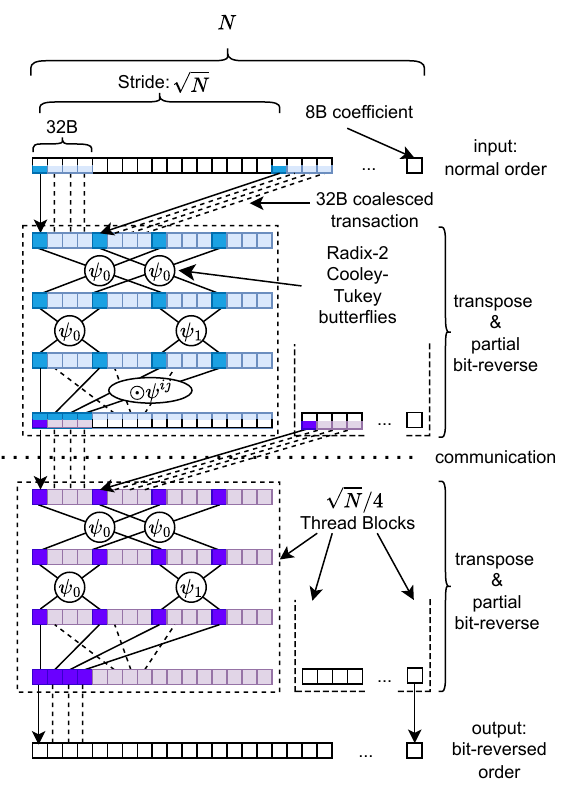}
}
\caption{Data movement diagram of the Radix-2 2D/Hierarchical NTT algorithm implemented in FIDESLib.}
\label{fig:NTT}
\end{figure}

As $N \in 2^{{13, \dots, 17}}$ for typical parameter sets, a single limb ($64-512$KB) does not fit into the shared memory of a Streaming Multiprocessor (SM). Therefore, the algorithm must rely on global memory accesses for communication. To minimize these accesses, FIDESlib employs a Hierarchical/2D NTT scheme. This approach reduces the NTT computation to just four memory accesses per element, significantly fewer than a naive 1D NTT implementation. Figure~\ref{fig:NTT} provides an overview of the data movement involved in our implementation.
As accesses to the coefficient vector must be performed with a stride, each thread block performs computation on four $\sqrt{N}$-sized sub-FFTs. This strategy leverages the memory block size of the GPU architecture. Additionally, performing multiple FFTs within a single thread block helps amortize the cost of loading the precomputed twiddle values ($\psi_i$) into shared memory.

To reduce L2 load bandwidth usage, the required full vector scaling steps compute twiddle factors ``on the fly'' using some of the input from the already loaded twiddle factor vector. By precomputing the twiddle factors, corresponding Shoup constants are also precomputed, enabling Shoup modular multiplication to speed up the NTT butterflies.

The iNTT is implemented with Gentleman-Sande butterflies instead of Cooley-Tukey butterflies, which take a bit-reversed evaluation polynomial and output a normal-ordered coefficient polynomial. This approach eliminates the need for explicit bit-reversal steps.

\subsubsection{Operation Fusions}
A common technique employed to reduce the memory bandwidth requirements of CKKS is kernel fusion. As the name suggests, several mathematical steps are computed within a single kernel, which reduces the frequency of data reads from and writes to global memory.

Before showcasing some instances of operation fusion employed by FIDESlib, we will discuss the implementation. The main drawback of implementing operation fusion is the combinatorial growth of possible kernel fusion combinations. The primary kernel fusion approach involves appending element-wise operations to the more compute-intensive (i)NTT kernels. To facilitate this, our (i)NTT kernels use a template to include kernel fusions without duplicating any code.

\begin{itemize}
    \item \textbf{Rescale fusion}: The NTT kernel starts with a fused $SwitchModulo$ operation and ends with the combined sequence: $q_l^{-1}(x^{(i)} - NTT(SwitchModulo(x^{(l)})))$, avoiding up to six memory operations per element.
    \item \textbf{ModDown fusion}: Similarly, the NTT kernel performs the $\mathcal{P}^{-1}(x^{(i)} - NTT(x'^{(i)}))$ sequence, saving up to four memory operations per element.
    \item \textbf{HMult fusion}: The outputs $\{\mathcal{P}*(c_0^{(i)}\odot c_0'^{(i)})*ksk_{0,d_i}^{(i)}, \mathcal{P}*(c_0^{(i)}\odot c_1'^{(i)} + c_1^{(i)} \odot c_0'^{(i)}) \odot ksk_{1,d_i}^{(i)}, INTT(c_1^{(i)}\odot c_1'^{(i)})\}$ are produced by the iNTT kernels before ModUp saving up to 10 memory operations per element. After the base conversion step, the NTT kernels generate $\{NTT(x'^{(i)})*ksk_{0,\beta}^{(i)} , NTT(x'^{(i)})*ksk_{1,\beta}^{(i)}\}$ as a result, saving up to three memory operations per element.
    \item \textbf{Dot product fusion}: The pattern $\sum_{i = 0}^{n - 1}{a \odot b}$ can be optimized from $6n-3$ memory operations per output element to $2n+1$ by computing everything in a single kernel. Similarly, a weighed sum can be reduced from $4n-2$ down to $n+1$ memory operations per element.  
\end{itemize}

\subsubsection{Hoisted rotation}
When multiple rotations of a single ciphertext are required, the ModUp operation is performed once before applying key switching with different keys to its output \cite{hs2018bts}. This optimization reduces the overall runtime to obtain the rotated ciphertexts. FIDESlib implements this routine, which is especially useful for accelerating Bootstrap execution.

\subsubsection{Bootstrapping}
FIDESlib adapts the CKKS bootstrapping procedure from OpenFHE for execution on GPU. The bootstrapping runtime consists of three main routines: CoeffToSlot, ApproxModEval, and SlotToCoeff \cite{chkks2018}.

FIDESlib adapts the ApproxModEval routine directly from OpenFHE \cite{bossuat2022}. It implements Chebyshev cosine approximation using the BSGS (Baby-Step Giant-Step) algorithm \cite{hs2018bts} and the Paterson-Stockmeyer method, followed by several iterations of the double angle formula to simulate a larger approximation range at the cost of increased level usage\cite{hk2020}.

For the CoeffToSlot and SlotToCoeffs steps, FIDESlib generalizes them into a single routine. These steps perform homomorphic encoding and decoding through the homomorphic application of a Discrete Fourier Transform (DFT). To improve efficiency at the cost of increasing multiplicative levels, the DFT plaintext matrix is split into several block matrices with higher sparsity \cite{chen2019improved}. Each ciphertext-vector times plaintext-matrix multiplication is then performed using a BSGS algorithm \cite{bossuat2021}, which reduces the number of required rotations and leverages the hoisted rotation optimization. Unlike OpenFHE, FIDESlib does not perform ModDown hoisting, this reduces the size of precomputed plaintexts and minimizes device memory usage.

\section{Evaluation}
\label{sec:evaluation}

\subsection{Methodology}
\label{subsec:methodology}

\begin{table*}[t!]
\centering
\begin{tabular}{l||ccccc|cc} % Updated column format to include vertical lines
\toprule
\textbf{Compute Platform}         & \textbf{Frequency} & \textbf{CPU Cores or SMs} & \textbf{32b INT TOPS} & \textbf{Private Data Cache} & \textbf{Shared Cache} & \textbf{DRAM Size} & \textbf{Bandwidth}  \\ 
\hline
CPU: Ryzen 9 7900 & $3.70$ GHz                        & $12$    & $2.13$            & $1056$ KB & $64$ MB         & $64$ GB & $81$ GB/s \\
\hline
GPU: RTX 4060 Ti & $2.31$ GHz                        & $34$    & $11.03$            & $128$ KB & $32$ MB         & $16$ GB & $288$ GB/s \\
GPU: RTX A4500   & $1.05$ GHz                       & $56$    & $11.83$            & $128$ KB & $6$ MB          & $20$ GB & $640$ GB/s \\
GPU: V100   & $1.25$ GHz                      & $80$    & $14.13$          & $128$ KB & $6$ MB         & $16$ GB & $897$ GB/s \\
GPU: RTX 4090    & $2.24$ GHz                      & $128$   & $41.29$          & $128$ KB & $72$ MB         & $24$ GB & $1$ TB/s \\
\bottomrule
\end{tabular}
\caption{Specifications of the CPU and GPU compute platforms used in our experiments.}
\label{tab:gpu_specs}
\end{table*}

% \begin{itemize}
%     \item ? - Validation of our implementation (OpenFHE comparison)
%     \item ? - Testbed (CPU used for executing OpenFHE and HEXL + GPU archs (ref Table III) + libraries)
%     \item ? - Benchmarks and microbenchmarks
    
% \end{itemize}

We have used Google Test to perform unit tests, and also integration tests with OpenFHE, developing a test suite that has been executed successfully using the following NVIDIA GPUs: 4060Ti, A4500, V100, and 4090 (details in Table~\ref{tab:gpu_specs}).
We compare the efficiency of FIDESlib with OpenFHE (baseline), an AVX-512 optimized OpenFHE implementation leveraging Intel's HEXL~\cite{boemer2021intelhexlacceleratinghomomorphic}, and the current leading open-source GPU implementation, Phantom.

In order to comprehensively evaluate all possible scenarios, all tests are executed for different CKKS parameter sets ($[N, L, \Delta, dnum]$). Tests for specific operations such as CiphertextSquaring are further specified with additional parameters, such as limb-batching configuration or ciphertext level.
For the performance evaluation of FIDESlib, we opted for Google Benchmark, a benchmarking tool that offers a similar interface to Google Test and allows for easy customization to collect a wide range of performance metrics. It also allows for the parametrization of benchmarks, enabling us to easily obtain performance information for FIDESlib using different parameter sets on various GPUs.

\subsection{Results}
\label{subsec:results}

We evaluate the performance of every public API operation in FIDESlib, as well as some internal operations, to obtain comprehensive data on all implemented functionality.
Unless otherwise specified, $[N, L, \Delta, dnum] := [2^{16}, 29, 59, 4]$.
%, which allows for bootstrapping.

The main performance results are presented in Table~\ref{tab:Ops_eval}. We show FIDESlib's operations execution time for a single maximum level ciphertext ($\ell$=29). We compare against a baseline single-threaded OpenFHE installation, multi-threaded and HEXL enabled OpenFHE, and the Phantom FHE open-source GPU library. The CPU tests are run on an AMD Ryzen 9 7900 (12-core, SMT enabled, AVX-512 enabled) with DDR5 memory at 5200 MT/s. All GPU tests are run on the same Nvidia RTX 4090 system. As we can see, FIDESlib achieves the best performance on every operation. Notably, HMult is more than $100\times$ faster than a multi-threaded CPU implementation, while Rescale is more than $30\times$ faster.

\begin{table*}[h]
\centering
\tiny
\resizebox{2\columnwidth}{!}{
\begin{tabular}{r||r|rr|rr|rr}
\toprule
& \begin{tabular}{c}
    OpenFHE \\
    (Baseline)
\end{tabular}
& \multicolumn{2}{c|}{\begin{tabular}{c}
    OpenFHE \\
    (Intel HEXL. 24 threads)
\end{tabular}}
& \multicolumn{2}{c|}{\begin{tabular}{c}
    Phantom \\
    (RTX 4090)
\end{tabular}}
& \multicolumn{2}{c}{\begin{tabular}{c}
    \textbf{FIDESlib} \\
    (RTX 4090)
\end{tabular}} \\
%\midrule
%Operation & Time & Time & Speedup & Time & Speedup & Time & Speedup \\
\hline
\textbf{ScalarAdd} & 1.28 \(\mathrm{ms}\) 
& 106.00 \(\mathrm{\mu s}\) & 12.06\(\times\) 
& \multicolumn{2}{c|}{N/A} 
& \textbf{16.63} \(\mathrm{\mu s}\) & 76.89\(\times\)\\
\textbf{PtAdd} & 5.26 \(\mathrm{ms}\) 
& 5.80 \(\mathrm{ms}\) & 0.90\(\times\) 
& 20.64 \(\mathrm{\mu s}\) & 254.87\(\times\) 
& \textbf{17.79} \(\mathrm{\mu s}\) & 295.65\(\times\)\\
\textbf{HAdd} & 7.84 \(\mathrm{ms}\) 
& 8.39 \(\mathrm{ms}\) & 0.93\(\times\) 
& 82.66 \(\mathrm{\mu s}\) & 94.86\(\times\) 
& \textbf{50.70} \(\mathrm{\mu s}\) & 154.67\(\times\)\\
\textbf{ScalarMult} & 4.34 \(\mathrm{ms}\) 
& 225.00 \(\mathrm{\mu s}\) & 19.29\(\times\) 
& \multicolumn{2}{c|}{N/A}
& \textbf{44.15} \(\mathrm{\mu s}\) & 98.32\(\times\)\\
\textbf{PtMult} & 10.14 \(\mathrm{ms}\) 
& 5.32 \(\mathrm{ms}\) & 1.90\(\times\) 
& 31.91 \(\mathrm{\mu s}\) & 317.25\(\times\) 
& \textbf{21.74} \(\mathrm{\mu s}\) & 465.72\(\times\)\\
\textbf{Rescale} & 50.80 \(\mathrm{ms}\) 
& 4.92 \(\mathrm{ms}\) & 10.31\(\times\) 
& 224.58 \(\mathrm{\mu s}\) & 226.21\(\times\) 
& \textbf{156.11} \(\mathrm{\mu s}\) & 325.44\(\times\)\\
\textbf{HRotate}* & 370.71 \(\mathrm{ms}\) 
& 105.30 \(\mathrm{ms}\) & 3.52\(\times\) 
& 1.139 \(\mathrm{ms}\) & 325.60\(\times\) 
& \textbf{1.107} \(\mathrm{ms}\) & 334.90\(\times\)\\
\textbf{HMult} & 406.24 \(\mathrm{ms}\) 
& 151.58 \(\mathrm{ms}\) & 2.62\(\times\) 
& 1.220 \(\mathrm{ms}\) & 332.99\(\times\) 
& \textbf{1.084} \(\mathrm{ms}\) & 374.61\(\times\)\\
\bottomrule
\end{tabular}
}
\caption{Performance comparison of CKKS primitives. *Conjugate is implemented exactly the same as HRotate.}
\label{tab:Ops_eval}
\end{table*}

\begin{figure}[t!]
\centerline {
    \includegraphics[width=0.4\textwidth, clip=true, trim=0 0 0 0mm]{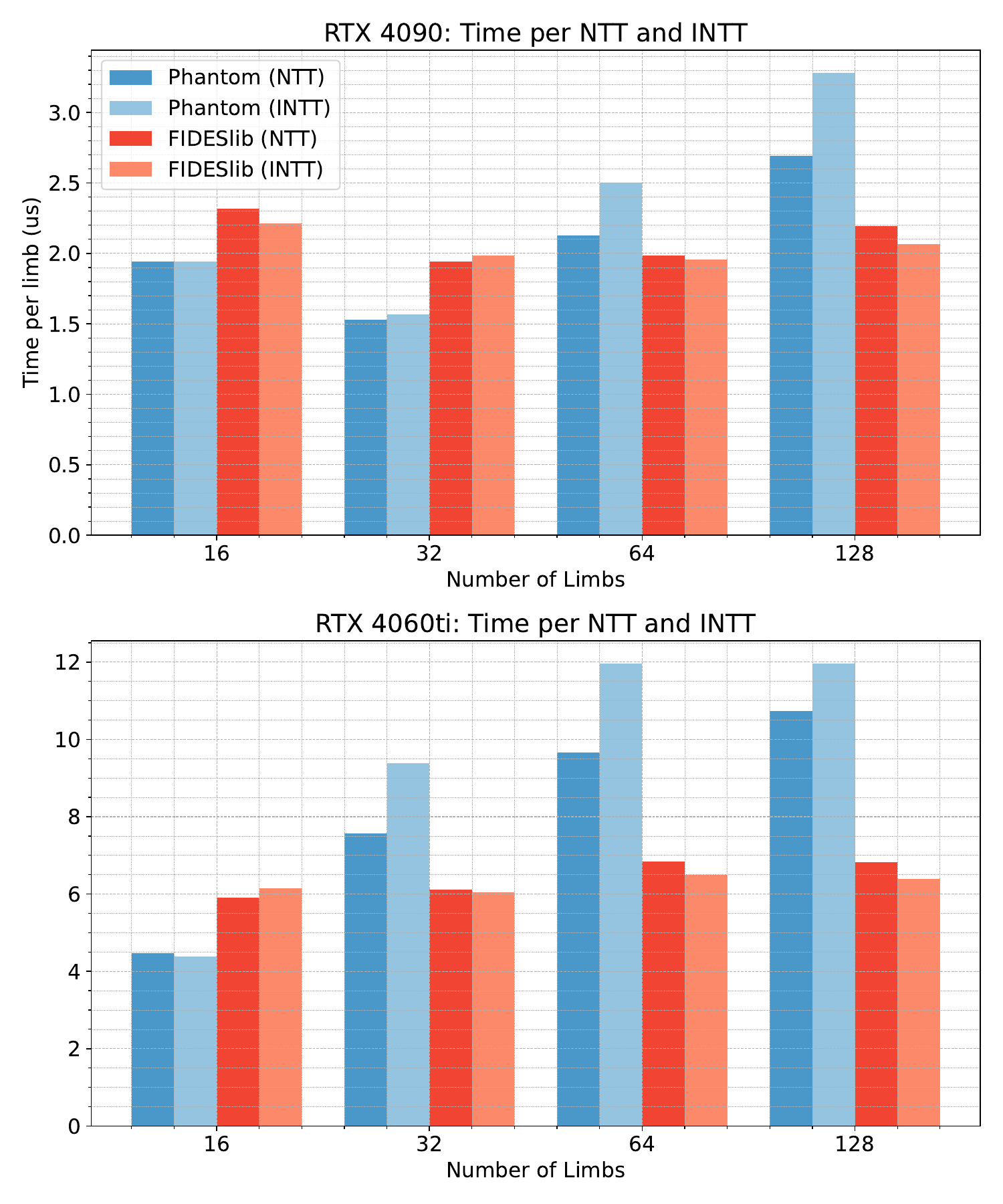}
    }
    \caption{(i)NTT performance comparison (lower is better).}
\label{fig:ntt_troughput}
\end{figure}

A performant and balanced (i)NTT implementation is crucial to achieve fast and reliable FHE performance on a variety of platforms. Figure~\ref{fig:ntt_troughput} shows a comparison of the time per limb on a range of limb working sets, in terms of size. For context, a ModUp operation involves computing up to $(dnum + 1)*L$ (around 150 for bootstrappable parameters) (i)NTT operations. Our implementation shows greater memory bandwidth efficiency as the working set grows.

\begin{figure}[t!]
\centerline {
    \includegraphics[width=0.4\textwidth, clip=true, trim=0 0 0 0mm]{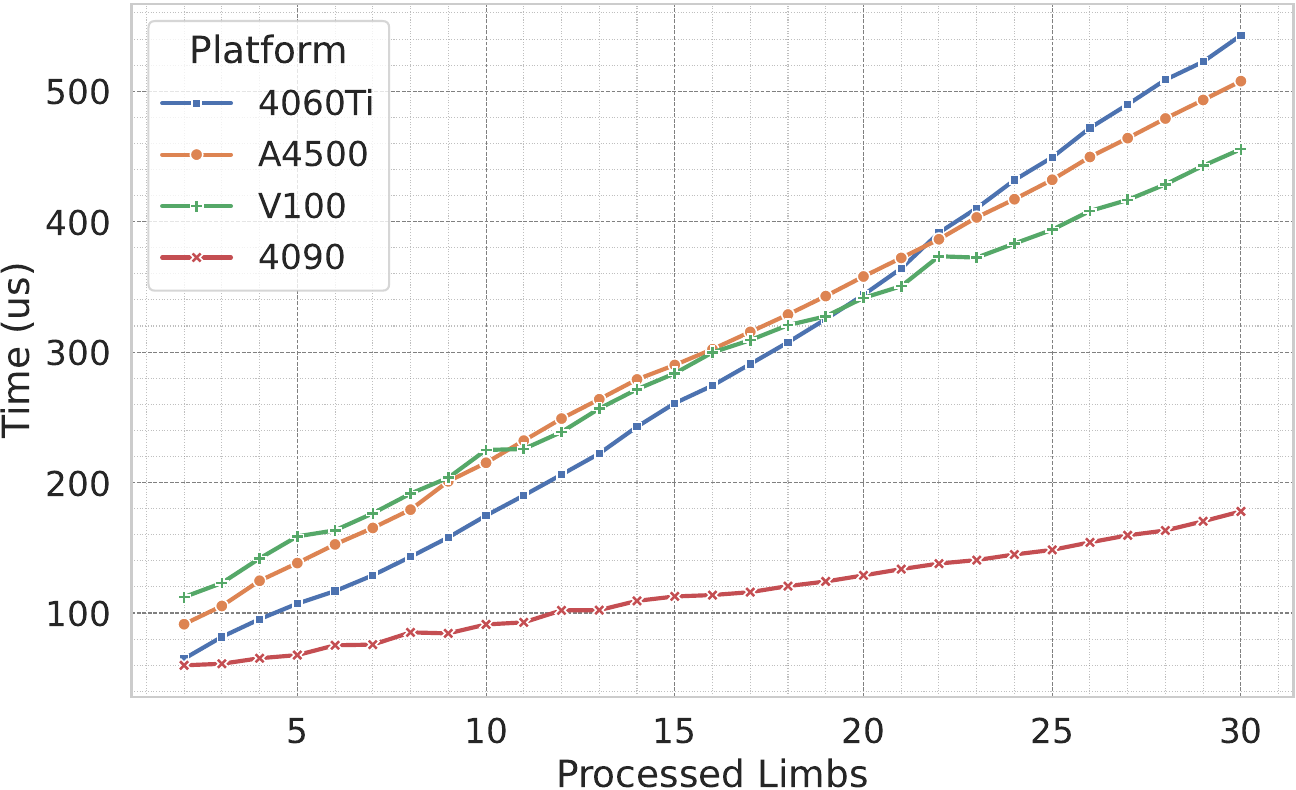}
    }
    \caption{$PtMult$+$Rescale$. Best limb batch.}
\label{fig:rescale_tp_best_batch}
\end{figure}

Since plaintext multiplication followed by a rescaling operation is a common pattern in CKKS, we evaluate this sequence in Figure~\ref{fig:rescale_tp_best_batch}. As observed, the time complexity is nearly linear with the number of RNS limbs involved. Some slight non-linearity is observed with the RTX 4060 Ti GPU, as the working set size starts to fit into its 32-MB L2 cache below the 20-limb data point. 

\begin{figure}[t!]
\centerline {
    \includegraphics[width=0.4\textwidth, clip=true, trim=0 0 0 0mm]{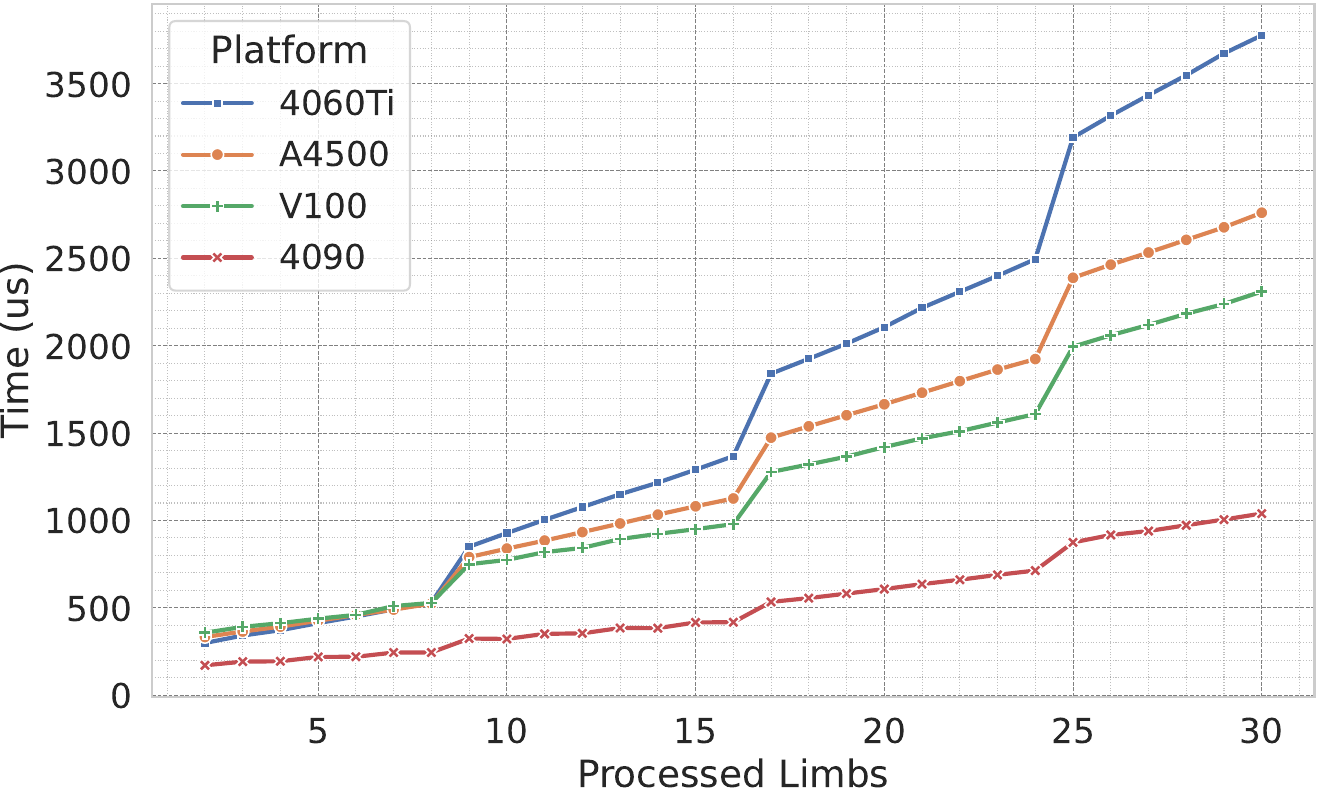}
    }
    \caption{$HMult$. Best limb batch.}
\label{fig:ciphertext_mult_tp_best_batch}
\end{figure}

Figure ~\ref{fig:ciphertext_mult_tp_best_batch} illustrates the performance of the HMult operation across various architectures. 
The limb batch parameter employed on each platform is the one that yielded the highest performance.
We showcase how performance varies depending on the amount of used levels of the ciphertext for a given parameter set. The Hybrid-Key-Switching technique causes a noticeable speedup each time a ``digit'' is dropped (as ciphertext multiplicative levels are used). 
%. 

% \begin{figure}[htbp]
% \centerline {
%     \includegraphics[width=0.5\textwidth, clip=true, trim=53 18 70 14mm]{figures/plots/CiphertextMultiplication-EPS-MeanConfidence.png}
%     }
%     \caption{Ciphertext Multiplication. Processed elements per second. Batch-wise average (95\% confidence).}
% \label{fig:ciphertext_mult_eps_mean_confidence}
% \end{figure}

% To provide a comprehensive performance evaluation, it is essential to demonstrate the impact of suboptimal batch sizes on different architectures. Figure~\ref{fig:ciphertext_mult_eps_mean_confidence} presents the mean and 95\% confidence interval of Elements Processed per Second metric. Intriguingly, for GPUs like the A4500, the performance variation across different batch sizes is minimal.

\begin{figure}[t!]
\centerline {
    \includegraphics[width=0.4\textwidth, clip=true, trim=0 0 0 0mm]{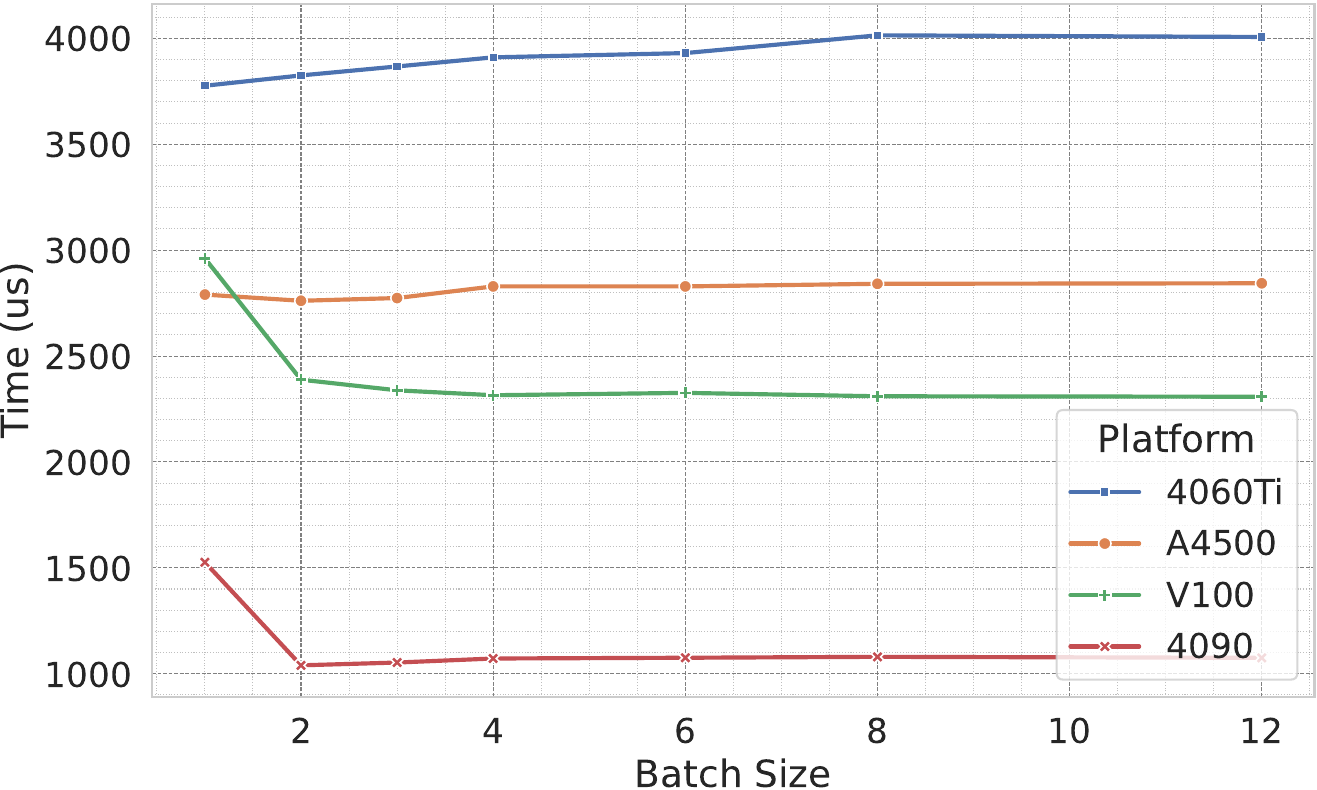}
    }
    \caption{$HMult$ with maximum levels. Limb batch comparison.}
\label{fig:ciphertext_mult_tp_batch_perf}
\end{figure}

Figure~\ref{fig:ciphertext_mult_tp_batch_perf} highlights the impact of the limb batch parameter on the performance of the HMult operation. As GPU throughput increases, the best performing configuration is found at a higher limb batch.

\begin{figure}[t!]
\centerline {
    \includegraphics[width=0.4\textwidth, clip=true, trim=0 0 0 0mm]{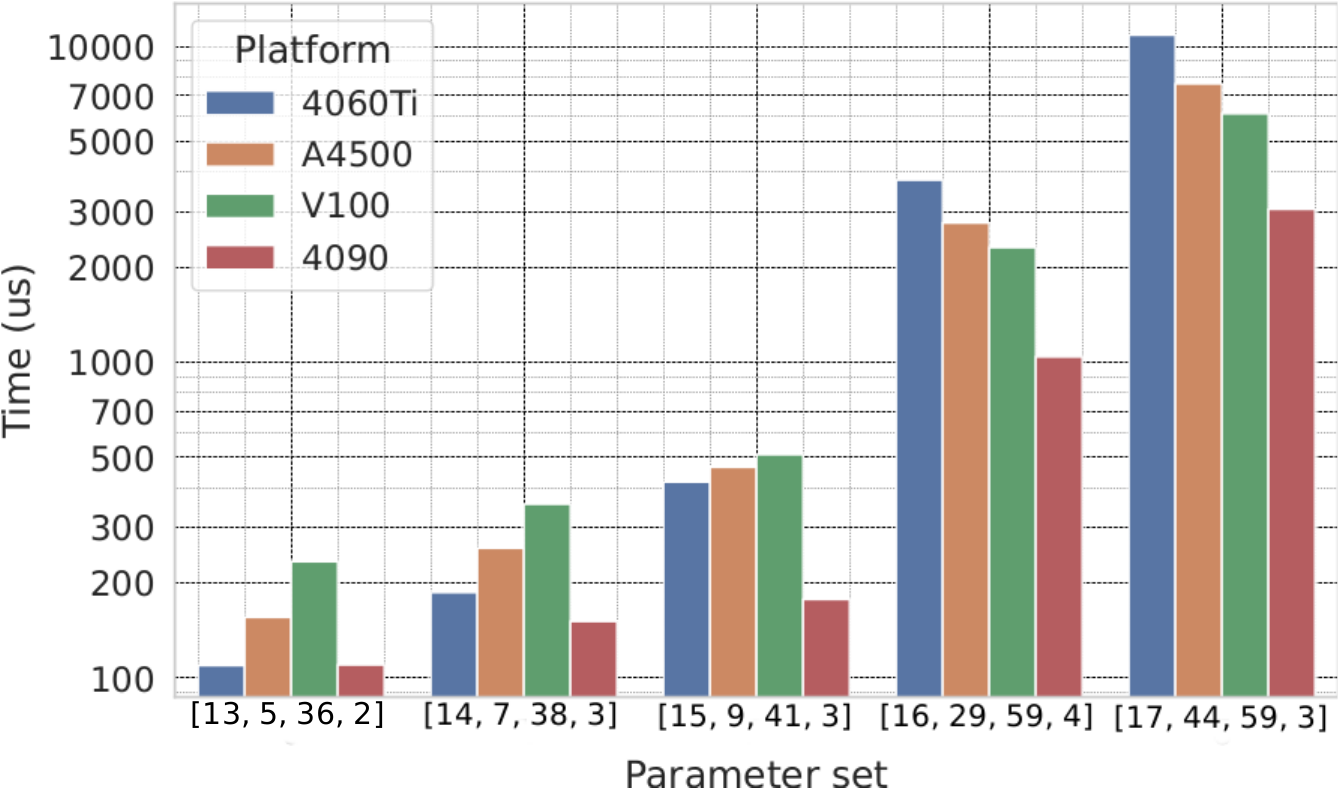}
    }
    \caption{$HMult$ with maximum levels. Parameter set $[log(N), L, \Delta, dnum]$ comparison.}
\label{fig:ciphertext_mult_tp_param_comparisons}
\end{figure}

Performance of the HMult operation is shown for different parameter sets in Figure~\ref{fig:ciphertext_mult_tp_param_comparisons}, showcasing FIDESlib's versatility and robustness. On smaller parameter sets, the workload is small and it is bottlenecked by kernel latency, so higher frequency platforms (RTX 4060 ti, RTX 4090) see improved performance over throughput oriented designs (V100). Furthermore, as key-switching key sizes vary ($2.3$MB, $7.7$MB, $20$MB, $152$MB, and $360$MB, respectively), some devices benefit from higher L2 cache hit-rate at some parameter levels, improving performance. 

Bootstrapping performance and throughput, compared to our baseline OpenFHE implementation and the Intel-optimized OpenFHE HEXL version, is presented in Table~\ref{tab:Bootstrap}. The execution times are comparable as the achieved message precision is equal across all implementations. Execution time increases with the number of slots (encoded values), though is amortized over time as we increase the number of slots. Performance is over $70\times$ faster than HEXL-enabled OpenFHE.

\begin{table}[t!]
\large
\centering
\resizebox{1\columnwidth}{!}{
\begin{tabular}{c|c||r|c|c|c}
\toprule
Slots & Levels  & \begin{tabular}{c}
    OpenFHE
\end{tabular}
& 
\multicolumn{2}{c|}{\begin{tabular}{c}
    HEXL-24 threads
\end{tabular}} & \begin{tabular}{c}
    \textbf{FIDESlib}-RTX 4090
\end{tabular} \\
\hline
\multirow{2}{*}{$64$} & \multirow{2}{*}{$13$}  & $18\;224\;T$ & \multicolumn{2}{c|}{$5\;204$ ($3.50\times$)} & \textbf{73.5} ($248\times$) \\
                        &                      & $21\;904\;A$ & \multicolumn{2}{c|}{$6\;255$} & \textbf{88.3} \\
\hline
\multirow{2}{*}{$512$} & \multirow{2}{*}{$11$} & $18\;268\;T$ & \multicolumn{2}{c|}{$7\;781$ ($2.35\times$)} & \textbf{93.3} ($196\times$) \\
                        &                      & $3\;243\;A$ & \multicolumn{2}{c|}{$1\;381$} & \textbf{16.6} \\
\hline
\multirow{2}{*}{$16384$} & \multirow{2}{*}{$9$} & $20\;079\;T$ & \multicolumn{2}{c|}{$9\;281$ ($2.16\times$)} & \textbf{112} ($179\times$) \\
                         &                      & $136\;A$ & \multicolumn{2}{c|}{$62.9$} & \textbf{0.761} \\
\hline
\multirow{2}{*}{$32768$} & \multirow{2}{*}{$9$}  & $28\;635\;T$ & \multicolumn{2}{c|}{$12\;185$ ($2.35\times$)} & \textbf{146} ($196\times$) \\
                         &                       & $97.1\;A$ & \multicolumn{2}{c|}{$41.3$} & \textbf{0.496} \\
\bottomrule
\end{tabular}
}
\caption{Bootstrapping performance. $Slots$=Number of encoded values; $Levels$=Levels remaining after bootstrapping; $T$=Time in ms; $A$=Amortized time in $\frac{\mu s}{\text{slots}\cdot\text{levels}}$.}
\label{tab:Bootstrap}
\end{table}

\begin{table}[t!]
    \large
\centering
\resizebox{\columnwidth}{!}{
    \begin{tabular}{c||c|c|c}
    \toprule
       \textbf{Time (ms)} &  \textbf{OpenFHE} & \textbf{HEXL - 24 threads} & \textbf{FIDESlib - RTX 4090}\\
         \hline
       Iteration  & 1,555 & 448 (3.47$\times$)& \textbf{23} (67.61$\times$)\\
       \hline
       Iteration + Bootstrap & 16,233 & 7,233 (2.24$\times$) & \textbf{169} (96.05$\times$)\\
       \bottomrule
    \end{tabular}
    }
    \caption{LR performance. Time in ms (and speedup). }
    \label{tab:lr_perf}
\end{table}

%4060Ti
%- FIDES: 1st it: NoBoot 118ms Boot 382ms
%         Hot: NoBoot 31ms   Boot 292ms
%Base
%- OpenFHE: 1st it: NoBoot 2615ms Boot 21481ms
%           Hot: NoBoot 2754ms Boot 21632ms
%HEXL
%- OpenFHE: 1st it: NoBoot 2912ms Boot 9655ms
%           Hot: NoBoot 720ms Boot 7155ms
%4090
%- FIDES: 1st it: NoBoot 72ms Boot 244ms
%         Hot: NoBoot: 30ms Boot 195ms
%A4500
%- FIDES: 1st it: NoBoot 112ms Boot 338ms
%         Hot: NoBooot: 33ms Boot: 253ms

To evaluate our library's robustness and performance under realistic workloads, we implemented a modified Logistic Regression (LR) algorithm based on the approach in \cite{lr_fhe}. We trained the model on a dataset of 45,000 loan eligibility samples using mini-batch standard gradient descent with ciphertexts containing 1,024 samples. Although each data sample had 25 parameters after encoding, we opted to align our data to the next power of two boundary, 32, to optimize the rotations needed on each iteration. Table~\ref{tab:lr_perf} shows the performance comparison of the same training algorithm both on OpenFHE and FIDESlib. We bootstrap every LR iteration while employing parameters $[N, L, \Delta, dnum] := [2^{16}, 26, 59, 4]$.

\section{Related Work}
\label{sec:related}

\begin{table}[t!]
\resizebox{\columnwidth}{!}{
\begin{tabular}{c:c||c|c|c|c|c|c|c|c|c}
\toprule
\multicolumn{2}{c}{\textbf{\huge Features}} 
& \textbf{\huge \cite{heaan2022}} 
& \textbf{\huge \cite{heongpu2024}} 
& \textbf{\huge \cite{jung2021over100}} 
& \textbf{\huge \cite{troynova2023}} 
& \textbf{\huge \cite{yang2024phantom}} 
& \textbf{\huge \cite{kim2024cheddar}} 
%& \textbf{\huge \cite{openfhe}} 
%& \textbf{\huge \cite{wang2023hebooster}} 
& \textbf{\huge \cite{desilo2023liberatefhe}} 
& \textbf{\huge \cite{fan2023tensorfhe}}
& \textbf{\huge FIDESlib} \\ 
\hline
\multicolumn{2}{l||}{\huge Open Source} & & \huge \greencheck & & \huge \greencheck & \huge \greencheck & & \huge \greencheck & & \huge \greencheck\\ 
\hline
\multicolumn{2}{l||}{\huge Published} & & & \huge \greencheck & & \huge \greencheck & \huge \greencheck & & \huge \greencheck & \huge \greencheck\\ 
\hline
\multicolumn{2}{l||}{\huge Bootstrapping} & \huge \greencheck & & \huge \greencheck & & & \huge \greencheck & & \huge \greencheck & \huge \greencheck\\
\hline
\multicolumn{2}{l||}{\huge OpenFHE Inter.} & & & & & & & & & \huge \greencheck\\ 
\hline
\multicolumn{2}{l||}{\huge Benchmarks}                  &                   &                   & \huge \greencheck &                   &                   &     \huge \greencheck&                   & \huge \greencheck & \LARGE $LR$\\      
\multicolumn{2}{l||}{\huge Microbench.}    &                   & \huge \greencheck & \huge \greencheck & \huge \greencheck & \huge \greencheck & \huge \greencheck &                   & \huge \greencheck & \huge \greencheck\\
\hline
\multicolumn{2}{l||}{\huge Unit Tests}                      &                   & \huge \greencheck &                   &                   \huge \greencheck &   &                   &                   &                   & \huge \greencheck\\
\multicolumn{2}{l||}{\huge Integration Tests}        &                                       &                   &                   &                   &                   &                   &                   &                   & \huge \greencheck\\
\hline
\multicolumn{2}{l||}{\huge Multi-GPU}    &                   &                   &                   & \huge \greencheck &                   &                   & \huge \greencheck &                   & \LARGE $WIP$\\
\bottomrule
\end{tabular}
}
\caption{Qualitative comparison of GPU CKKS libraries. $LR$=Logistic Regression; $WIP$=Work in progress.}
\label{tab:lib_comparisons}
\end{table}

Table~\ref{tab:lib_comparisons} shows a qualitative comparison of the state-of-the-art GPU-based CKKS libraries, while consider a range  of features. 
Phantom~\cite{yang2024phantom} is the most efficient open-source CKKS library for a GPU backend up to date, and is why we have used it as our GPU baseline for performance analysis in the previous section. However, it lacks several CKKS operations such as ScalarAdd, ScalarMult, HSquare, and bootstrapping. 
Libraries such as HEaaN~\cite{heaan2022}, Over100x~\cite{jung2021over100},  Cheddar~\cite{kim2024cheddar}, and TensorFHE~\cite{fan2023tensorfhe} offer feature-complete CKKS implementations, including bootstrapping. Nonetheless, the lack of publicly available source code hinders reproducibility and further research. Older versions of HEaaN and Over100x are publicly available; however, to the best of our knowledge, the open-source HEaaN code only supports a CPU backend, and the Over100x GPU backend is incomplete, preventing full reproduction of the reported experiments and limiting support for other CKKS parameter settings.

%This prevents full reproduction of the experiments reported in the publication and restricts the available parameter settings, as compared to the comprehensive evaluation methodology used in Section~\ref{sec:evaluation}.
In contrast, FIDESlib is the first fully-fledged open-source CKKS library designed for GPUs, delivering better performance than Phantom. FIDESlib is designed to be fully compatible with OpenFHE~\cite{openfhe}, the widely trusted industry standard for FHE. FIDESlib is the only library that ensures complete interoperability with OpenFHE, accelerating server-side CKKS operations, while offloading encoding and encryption tasks to OpenFHE-based clients.

% The double-scaling technique, promising performance-wise when combined with $32$-bit data types, is not supported by OpenFHE yet.

Robust testing of library functionality is lacking in this field; only HEonGPU~\cite{heongpu2024}, Troy-Nova\cite{troynova2023} (both are available for download, but have not been published in peer-reviewed venues), and OpenFHE~\cite{openfhe} offer some unit tests.
In contrast, FIDESlib features an extensive suite of unit tests that, combined with our extensive set of microbenchmarks, and our Logistic Regression benchmark, thoroughly validate all implemented functionality. Additionally, our new library represents, to our knowledge, the first instance of validation enhancement using integration tests that compare FIDESlib's output with OpenFHE results. Moreover, leveraging Google Test and Google Benchmark makes implementing continuous integration in our library straightforward. 

Although Liberate-FHE\cite{desilo2023liberatefhe} and Troy-Nova (both unpublished, non-peer-reviewed libraries) support multi-GPU backends for CKKS, Troy-Nova imposes limitations, such as requiring manual transfer of objects between devices. Liberate-FHE does support multi-GPU by dividing the workloads at the RNS level, but their implementation is an order of magnitude slower than other state-of-the-art implementations. FIDESlib's architecture will allow for the implementation of a more scalable multi-GPU solution, built upon a more performant baseline.

\section{Conclusions and Future Work}
\label{sec:conclusions}
This paper introduces FIDESlib, the first open-source CKKS library with full functionality optimized for a GPU backend, including bootstrapping, and featuring robust security from OpenFHE’s client-side operations.
Compared to Phantom, the current best-performing alternative, FIDESlib delivers better scalability and performance across all shared operations on four tested GPU platforms. 
For the most time-consuming and challenging bootstrapping operation, which is not supported in Phantom, FIDESlib achieves up to 227.8$\times$ and 74.4$\times$ speedups over OpenFHE and AVX-optimized OpenFHE, respectively.
FIDESlib is built with modern software practices, supporting extensibility and broad adoption. Our roadmap includes AMD GPU support via HIP~\cite{sun2022}, multi-GPU backends for NVIDIA and AMD, and porting more real-world MLaaS cloud-based workloads.

\section*{Acknowledgements}
The present work has been financed through the grant CNS2023-144241 funded by MICIU/AEI/10.13039/501100011033 and by the European Union NextGenerationEU/PRTR. Also in the context of the grant RYC2021-031966-I funded by MICIU/AEI/10.13039/501100011033 and by the European Union NextGenerationEU/PRTR. Finally, we thank the support by the grants NSF CNS 2312275, and 2312276, as well as, in part, the NSF IUCRC Center for Hardware and Embedded Systems Security and Trust (CHEST).

\bibliographystyle{IEEEtran}
\bibliography{references}

\end{document}